# The Migrating Embryo Model for Disk Evolution

Shantanu Basu[1] and Eduard I. Vorobyov[2,3]

[1] Department of Physics and Astronomy, Western University, London, Ontario, N6A 3K7, Canada
[2] University of Vienna, Institute of Astrophysics, Vienna, 1180, Austria
[3] Research Institute of Physics, Southern Federal University, Rostov-on-Don, 344090 Russia

**ABSTRACT**

A new view of disk evolution is emerging from self-consistent numerical simulation modeling of the formation of circumstellar disks from the direct collapse of prestellar cloud cores. This has implications for many aspects of star and planet formation, including the growth of dust and high-temperature processing of materials. A defining result is that the early evolution of a disk is crucially affected by the continuing mass loading from the core envelope, and is driven into recurrent phases of gravitational instability. Nonlinear spiral arms formed during these episodes fragment to form gaseous clumps in the disk. These clumps generally migrate inward due to gravitational torques arising from their interaction with a trailing spiral arm. Occasionally, a clump can open up a gap in the disk and settle into a stable orbit, revealing a direct pathway to the formation of companion stars, brown dwarfs, or giant planets. At other times, when multiple clumps are present, a low mass clump may even be ejected from the system, providing a pathway to the formation of free-floating brown dwarfs and giant planets in addition to low mass stars. Finally, it has been suggested that the inward migration of gaseous clumps can provide the proper conditions for the transport of high-temperature processed solids from the outer disk to the inner disk, and even possibly accelerate the formation of terrestrial planets in the inner disk. All of these features arising from clump formation and migration can be tied together conceptually in a Migrating Embryo model for disk evolution that can complement the well-known Core Accretion model for planet formation.

## 1. INTRODUCTION

The overall process of star formation can lead to the secondary outcomes of disk and planet formation. The early phase of disk formation and evolution holds the key to understanding stellar mass accumulation as well as sets the initial conditions for planet formation. The Core Accretion (CA) model (e.g., Pollack et al. 1996, Hubickyj et al. 2005) provides a successful conceptual framework to understand the formation of gas giant planets in a disk, beginning with the coagulation of planetesimals. However, work in recent years has shown that there is another compelling model for understanding disk evolution: the result that recurrent gravitational instability and fragmentation in a young disk (that may still be accumulating significant matter from its cloud core envelope) leads to the formation of gaseous, gravitationally-bound clumps that are then torqued inward so that they migrate toward merger with the central star (Vorobyov and Basu 2005, 2006, 2010b). These can trigger an FU Orionis type luminosity outburst. This result forms the basis of a new conceptual model, which we call the Migrating Embryo (ME) model. The ME model is more general than the idea of disk gravitational instability and fragmentation (GI), since it posits that clumps arising from disk fragmentation in the outer regions (tens to hundreds of AU from the primary object, due to thermodynamic constraints) generally migrate inwards until they are absorbed by the central object and/or sheared apart. Some clumps may be dispersed early on by tidal torques from spiral arms (Boley et al. 2010, Vorobyov and Basu 2010b). In some cases, a clump can be ejected from the system through many-body interactions (Basu and Vorobyov 2012), and in other cases a clump can clear a gap in the disk and settle into a wide orbit of several tens of AU radius (Vorobyov and Basu 2010a).

The destruction of clumps by tidal torques may in fact be as important for disk evolution as is their survival. Embryos that are driven into the disk inner regions may form solid cores in their interior via dust sedimentation or dense and compact atomic hydrogen cores via dissociation of molecular hydrogen. If one or both of those processes take place before the embryo is destroyed on its approach to the central star, a planet can emerge on orbit of order several AU or less (Nayakshin2010a; Boley et al. 2010).Since embryos are also sites of accelerated dust growth and high-temperature processing (Boss 1998; Boss et al.2002; Boley et al. 2010; Nayakshin 2010a, 2010b; Nayakshin et al. 2011), their in situ destruction by tidal torques may also release processed dust directly into the disk outer regions. Vorobyov (2011) has presented a model for the enrichment of embedded circumstellar disks with crystalline silicates formed in the depths of massive embryos via thermal annealing of amorphous dust grains. Part of the crystalline-silicate-bearing embryos are destroyed before they vaporize dust grains in their interiors or form solid cores via dust sedimentation, thus enriching the gas disk with crystalline silicates on radial distances from sub-AU to hundred-AU scales.

In the paper we review several of the above described aspects of the ME model, successively describing disk fragmentation, the burst mode of accretion, giant protoplanet formation, and the ejection of low mass or substellar objects. Section 7 gives an emphasized presentation on the production of crystalline silicates in the disk.

## 2. MODEL DESCRIPTION

The main concepts of our numerical approach are explained in detail in Vorobyov and Basu (2010b) and, for the reader's convenience, are briefly reviewed below. We start our numerical integration in the pre-stellar phase, which is characterized by a collapsing *starless* cloud core, continue into the star and disk formation phase, and terminate our simulations in the T Tauri phase, when most of the core has accreted onto the forming star/disk system. In the early evolution, the protostellar disk is not isolated but is exposed to intense mass loading from the envelope, an important prerequisite for realistic studies of the disk stability properties.
We introduce a "sink cell" at $r_{\rm sc}$= 6 AU and impose a free inflow inner boundary condition and free outflow outer boundary condition so that the matter is allowed to flow out of the computational domain but is prevented from flowing in. We monitor the gas surface density in the sink cell and when its value exceeds a critical value for the transition from isothermal to adiabatic evolution, we introduce a central point-mass star. In the subsequent evolution, 90% of the gas that crosses the inner boundary is assumed to land onto the central star plus the inner axisymmetric disk at $r$ <6 AU. The other 10% of the accreted gas is assumed to be carried away by protostellar outflows. Since we introduce the sink cell, no disk structure is resolved inward of 6 AU in our numerical hydrodynamics simulations.

We make use of the thin-disk approximation to compute the gravitational collapse of rotating, gravitationally unstable cloud cores. This approximation is an excellent means of calculating the evolution for many orbital periods and many model parameters and its justification is provided in Vorobyov and Basu (2010b). The basic equations of mass, momentum, and energy transport in the thin-disk approximation are

$$\frac{\partial \Sigma}{\partial t} + \nabla_p \cdot (\Sigma v_p) = 0,$$

$$\frac{\partial}{\partial t}(\Sigma \boldsymbol{v}_p) + [\boldsymbol{\nabla} \cdot (\Sigma \boldsymbol{v}_p \otimes \boldsymbol{v}_p)]_p = -\boldsymbol{\nabla}_p P + \Sigma \boldsymbol{g}_p + (\boldsymbol{\nabla} \cdot \boldsymbol{\Pi})_p,$$

$$\frac{\partial e}{\partial t} + \boldsymbol{\nabla}_p \cdot (e\,\boldsymbol{v}_p) = -P(\boldsymbol{\nabla}_p \cdot \boldsymbol{v}_p) - \Lambda + \Gamma + (\boldsymbol{\nabla} v)_{pp'}:\boldsymbol{\Pi}_{pp'},$$

where $\Sigma$ is the mass surface density, $\boldsymbol{v}_p$ is the velocity in the plane of the disk with $(r,\phi)$ components, $P$ is a vertically-integrated pressure, $\boldsymbol{g}_p$ is the gravitational field in the plane, $e$ is the internal energy per surface area, and $\boldsymbol{\nabla}_p$ is the gradient along the planar coordinates of the disk. A small amount of viscous transport is included using the α-prescription. The actual expressions used in polar coordinates for the viscous terms including the divergence of the stress tensor $(\boldsymbol{\nabla} \cdot \boldsymbol{\Pi})_p$, symmetrized velocity gradient tensor $\boldsymbol{\nabla} v$, viscous heating $(\boldsymbol{\nabla} v)_{pp'}:\boldsymbol{\Pi}_{pp'}$, and symmetric dyadic $\Sigma \boldsymbol{v}_p \otimes \boldsymbol{v}_p$ can be found in Vorobyov and Basu (2010b).

Initial cloud cores have surface densities $\Sigma$ and angular velocities $\Omega$ typical for a collapsing, axisymmetric, magnetically supercritical core:

$$\Sigma = \frac{r_0 \Sigma_0}{\sqrt{r_0^2 + r^2}},$$

$$\Omega = 2\Omega_0 \left(\frac{r_0}{r}\right)^2 \left[\sqrt{\left(1+\frac{r}{r_0}\right)^2} - 1\right],$$

(see Basu 1997) where $\Omega_0$ is the central angular velocity and $r_0$ is the radius of central near-constant-density plateau. The initial temperature in the core is 10 K and the radial velocity is set to zero. The core is gravitationally unstable and begins to collapse under its own gravity. Further details can be found in Vorobyov and Basu (2010b). We note that the above form of the column density is very similar to the integrated column density of a Bonnor–Ebert sphere (Dapp and Basu 2009).

## 3. DISK GRAVITATIONAL FRAGMENTATION

Figure 1 shows a series of images of the gas surface density in the inner 1000 AU at different times since the formation of the central star. The centrifugal disk can only form after the formation of a central point-mass potential (Basu and Mouschovias 1995), and in this model forms at a time $t = 0.08$ Myr after the start of the prestellar collapse. By $t = 0.13$ Myr, a well-developed spiral pattern and several dense clumps are clearly visible. The clumps are almost always located in the spiral arms, suggesting that they form via fragmentation of the densest and coldest arms. Most fragments, however, do not live long. They are driven into the disk inner regions and through the sink cell (and probably onto the star) but other fragments take their place. Some of them are massive enough to host mini-disks of their own. Typical fragment masses lie in a wide range from several Jovian masses to low- and intermediate-mass brown dwarfs. The mass spectrum of the fragments depends on the disk and cloud core properties and may vary from model to model.

The disk in this early phase of evolution is most certainly not in a steady-state. Regions that undergo fragmentation are ones that have become unstable according to the Toomre criterion, i.e., $Q = c_s \Omega / (\pi G \Sigma) < 1$. The cooling time of the gas is also typically less than the local dynamical time $\Omega^{-1}$ in these regions. Finally, fragmentation can be considered to be driven by the global criteria of a sufficiently large angular momentum in the system and sufficient mass infall from the core envelope onto the disk, so that the latter can be driven to develop regions that are Toomre unstable. A detailed discussion of these criteria is given in Vorobyov and Basu (2010b).

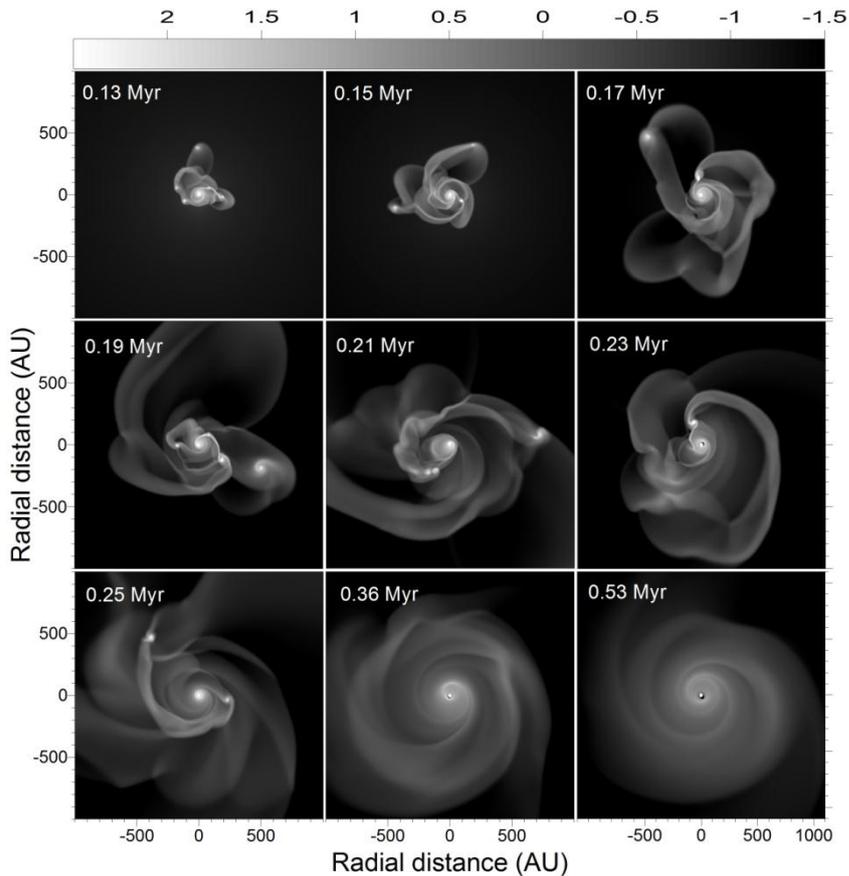

Fig. 1. Face view of the protostellar disk in model with mass $M_c = 0.7\ M_\odot$ and magnitude of the ratio of rotational energy to gravitational energy $\beta = 1.3*10^{-3}$. Shown is the integrated gas surface density (in g cm$^{-2}$) in the 2000 times 2000 AU box. Time elapsed since the formation of the central star (located at the coordinate origin) is indicated in each panel.

## 4. THE BURST MODE OF ACCRETION

Figure 2 shows, for the same model presented in Figure 1, the mass accretion rate onto the central sink and the inferred stellar luminosity if the mass accretes onto the star, both as a function of time elapsed since the beginning of collapse of the prestellar core.

In the prestellar phase, the accretion rate is negligible but quickly rises to $10^{-5} M_\odot$ yr$^{-1}$ when the gas volume density in the sink cell exceeds $10^{11}$ cm$^{-3}$ and a central stellar core begins to form at $t = 0.08$ Myr after the onset of collapse. The subsequent short period of near-constant accretion corresponds to the phase when the infalling envelope lands directly onto the forming star. A sharp drop in accretion rate follows shortly, which manifests the beginning of the disk formation phase. Subsequently, the infalling envelope accretes onto the forming disk rather than directly onto the star. This transient drop in accretion rate occurs due to the fact that the disk mass is initially too small to drive a substantial accretion rate onto the star either due to viscous or gravitational torques. As evolution proceeds, the disk accretes mass from the infalling envelope and a qualitatively new phase of mass accretion ensues, in which the accretion rate shows variability by several orders of magnitude. Short episodes of high-rate accretion (caused by the passage of disk fragments through the sink cell) are followed by longer periods of low-rate accretion (caused by a temporary disk expansion and stabilization). This highly variable accretion makes the star sporadically increase its *total* luminosity, as illustrated in the bottom panel of Figure 2. Several clear-cut luminosity outbursts with $L_{\rm accr}$ as high as 100 $L_\odot$ and many more weaker bursts are evident. The stronger bursts may represent FU Orionis-like eruptions (FUors), while weaker ones may manifest EX Lupi-like eruptions (EXors). FUors are young protostars which are believed to undergo accretion bursts during which the accretion rate rapidly increases from typically $10^{-7}$ $M_\odot$ yr$^{-1}$ to $10^{-4}$ $M_\odot$ yr$^{-1}$, and remains elevated over timescales of at least several decades (Hartmann and Kenyon 1996). EXors, a loosely defined class of pre-main-sequence stars, exhibit smaller accretion bursts on shorter timescales and may represent a later evolution stage of a protostar. We note that the exact time for the onset of the photospheric luminosity is rather uncertain and may shift to later times, which would result in the early luminosity bursts being considerably stronger in amplitude.

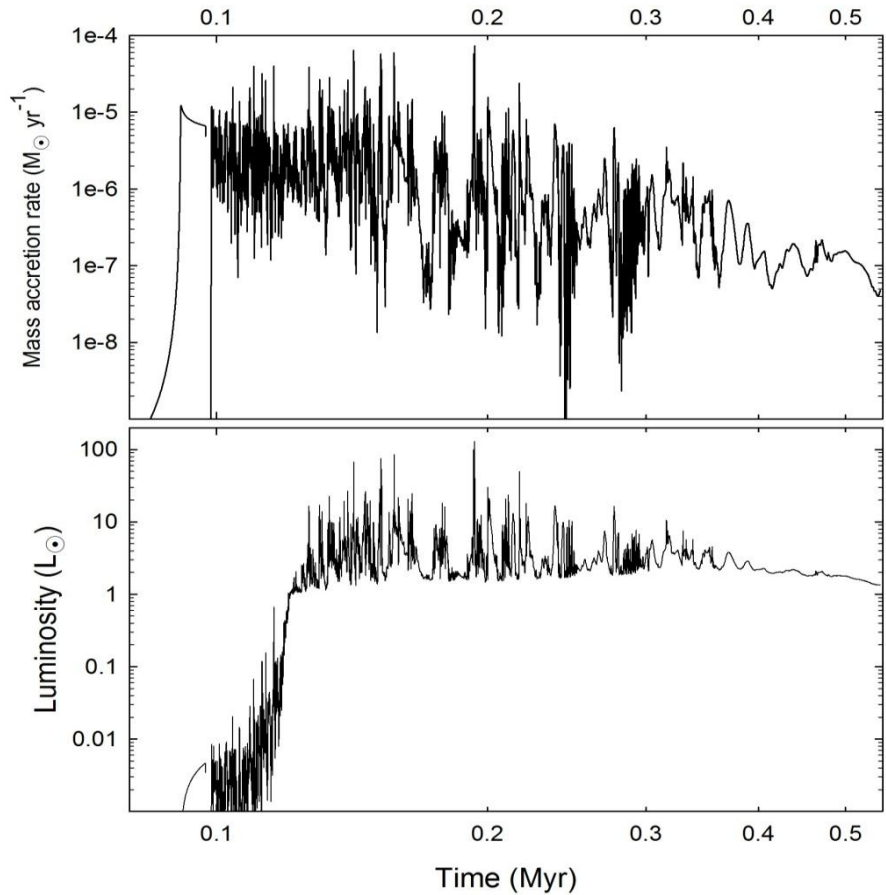

Fig. 2. (**Top.**) Mass accretion rate onto the protostar versus time in model with $M_c = 0.7$ $M_\odot$ and $\beta = 1.3*10^{-3}$. (**Bottom.**) Total (accretion plus photospheric) luminosity versus time in the same model.

## 5. SURVIVAL OF FRAGMENTS AND GIANT PLANET FORMATION

In Vorobyov and Basu (2010a), we presented a model of disk formation and evolution in which one of the fragments escaped fast inward migration and matured into a giant protoplanet (GPP) that was able to carve out a gap in the disk and settle into a stable orbit. This kind of outcome becomes increasingly probable as the mass and angular momentum content of the initial core is increased but is still quite rare and essentially stochastic in nature. Even for an almost identical set of initial parameters we may or may not see the formation of a GPP due to the fact that the disk evolution followed a slightly different pathway. In most cases, fragments do not mature into GPPs in wide orbits; they either migrate inward to the inner boundary or are dispersed in outer regions, possibly due to insufficient resolution. By studying a range of models with many realizations and multiple sets of initial parameters, we found that six out of 82 models formed GPPs on fixed orbits with final masses in the range of 5 – 10 $M_J$ and with orbit radii in the range of 25 – 200 AU. These models employ a barotropic relation between pressure and density, which makes a smooth transition from the early isothermal regime to a late adiabatic regime. So the treatment of thermal physics is more simplified in this model than in those presented in the previous and subsequent sections.

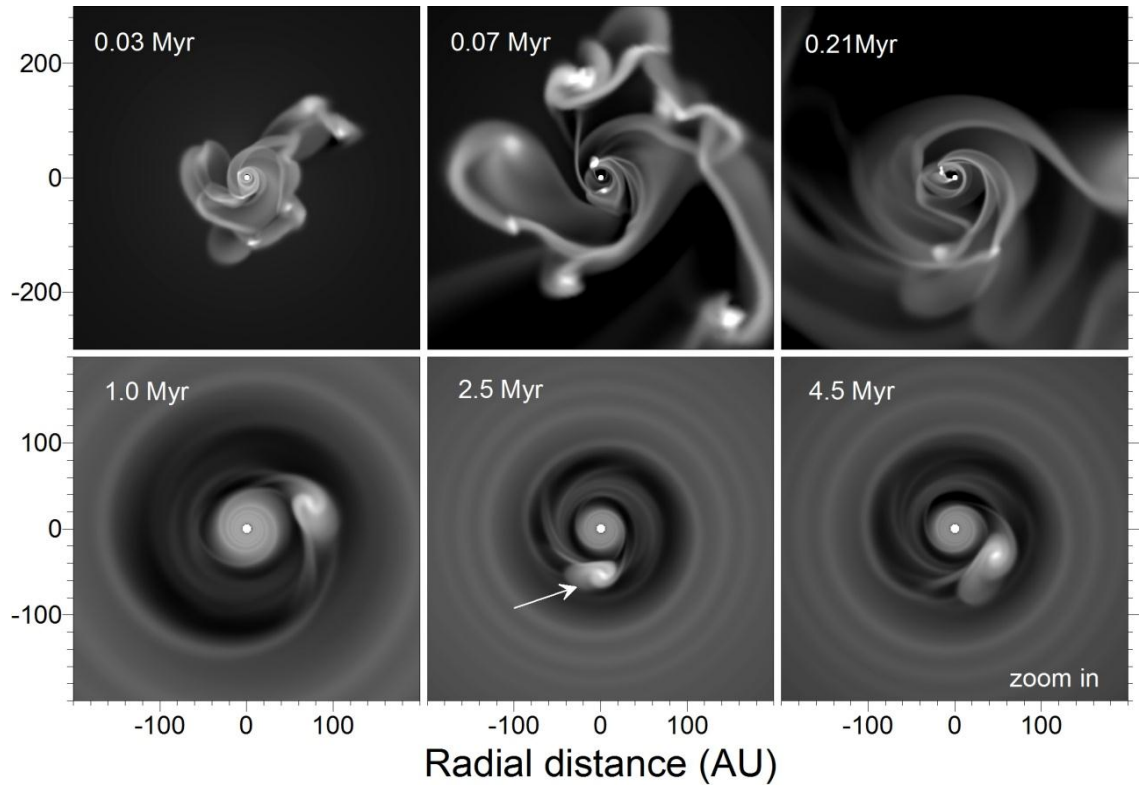

Fig. 3. Formation of a giant protoplanet with mass ≈ 5 $M_J$ and mean orbit distance ≈ 55 AU. Shown is the gas surface density (in g cm$^{-2}$) in the 600 times 600 AU box (top panels) and 400 times 400 AU box (bottom panels) at several consecutive times after the formation of the central star.

Figure 3 presents the gas surface density maps (in g cm$^{-2}$) in a model with initial core mass $M_c$ = 0.9 $M_\odot$ and ratio of rotational to gravitational energy magnitude β = 1.3 × 10$^{-2}$ at six evolution times after the formation of the disk. The rotation is counterclockwise (note that we zoom in at the bottom row). Several fragments condense in the outer parts of the spiral arms as early as 0.09 Myr after disk formation, but none of them have survived by the end of the embedded phase at 0.16 Myr when about 75% - 80% of the envelope has been accreted by the disk. They are all driven into the sink cell via a very efficient exchange of angular momentum with the spiral arms, possibly leading to multiple FU Orionis bursts (Vorobyov and Basu 2005, 2006) or forming giant and icy planets on very close orbits (Nayakshin 2010a, 2010b). The byproduct of these bursts is disk expansion due to the conservation of angular momentum. When the mass loading from the envelope diminishes and the burst phase ends, this expansion is followed by transient disk contraction, during which the gas surface density increases and several more fragments form in the disk's outermost regions ($t$ = 0.21 Myr). These fragments quickly migrate within the inner 100 AU.

The number of fragments formed during the early disk evolution varies from model to model and is usually between a few and few tens. The masses of the fragments lie in the giant planet and brown dwarf regime. By $t = 0.3$ Myr, only one fragment survives, which later opens a gap and evolves into a well-defined GPP possessing its own *counter-rotating* minidisk. Such counter-rotating minidisks are seen around many fragments. We believe that this effect is caused by the gravitational capture of some of the neighboring material, which receives a counter-rotating twist around the forming fragment due to differential rotation of the natal spiral arm.

## 6. EJECTION OF BROWN DWARFS AND VERY LOW MASS STARS

Recent work shows that as models increase along the sequence of increasing mass or angular momentum, there is also a tendency for a clump to be *ejected* from a disk due to interaction between multiple clumps and the primary object. This supports a new hybrid paradigm of brown dwarf (BD) and low mass star formation by disk fragmentation followed by clump ejection (rather than by ejection of finished low-mass stars/BDs). This scenario naturally accounts for the presence of disks around BDs and the relatively low ejection speeds in this scenario can account for the observed velocity dispersion and physical location of BDs relative to young stellar objects (YSOs). In principle, this mechanism could also eject protoplanetary mass fragments, leading to the formation of free-floating planets although we have not yet seen this in our numerical simulations. The total masses of ejected fragments (including circumfragemnt material) lie in the 0.05-0.34 $M_\odot$ limits, the ejection speeds – in the 0.38-1.2 km s$^{-1}$ limits, and the ejected times since the formation of the central protostar – 0.18-0.82 Myr. For more detailed statistics of the ejected fragments the reader is referred to Table~1 in Basu and Vorobyov (2012).

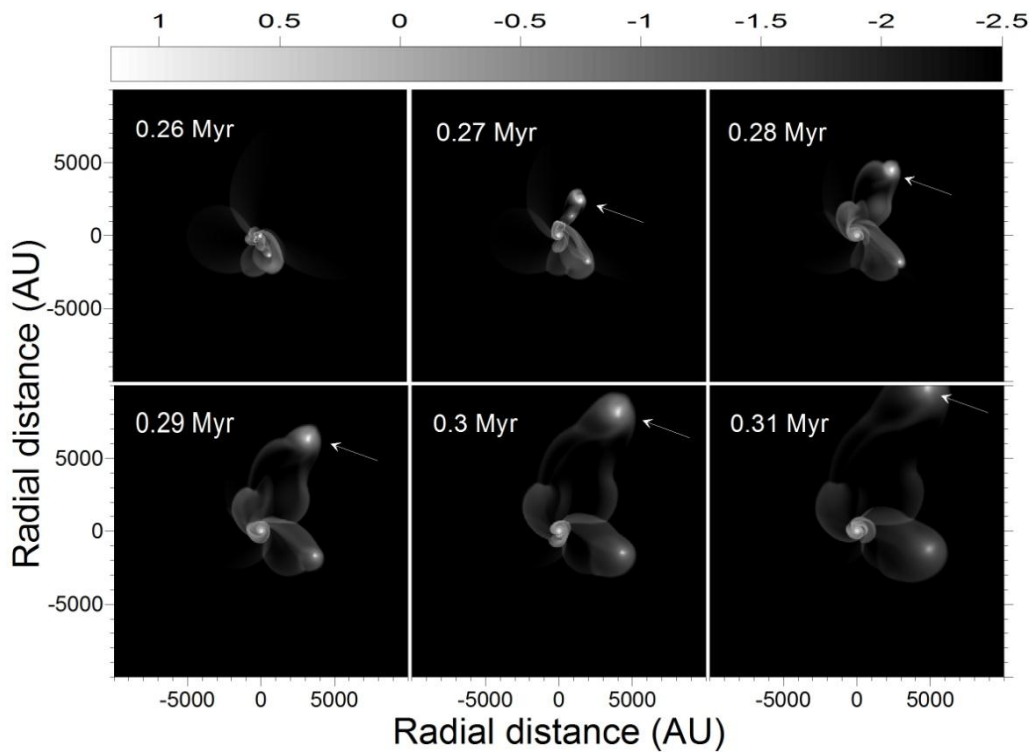

Fig. 4. Ejection of a proto-brown-dwarf embryo (shown by the arrows) from the protostellar disk (the inner region) due to many-body interaction. Shown is the gas surface density (in g cm$^{-2}$) in the 20000 times 20000 AU box at several consecutive times after the ejection instant at $t = 0.26$ Myr.

Figure 4 shows a sequence of column density images for a model at various times after the formation of a central object, and in a region of size 20,000 AU on each side. The circumstellar disk forms at $t = 0.01$ Myr and by $t = 0.05$ Myr it already undergoes fragmentation. Within the first 0.25 Myr, multiple fragments are formed in the relatively massive disk, at distances greater than about 50 AU and less than a few hundred AU. They are generally torqued inward through gravitational interaction with trailing spiral arms, as first found by Vorobyov and Basu (2005, 2006). Others located at large radii may eventually disperse. However, under sometime-favorable conditions, a clump within a multi-clump environment can be ejected through many-body interaction. Such an event begins at about 0.26 Myr. The ejection is also aided by the non-axisymmetric potential of the relatively massive disk. The velocity of the ejected clump (at the moment when it leaves the computational boundary at $r = 12,000$ AU) is about 0.9 km s$^{-1}$, which is a factor of three greater than the escape velocity $v_{\rm esc} = (2GM_{\rm sd}/r)^{1/2}$, where the total mass of the central star and its disk is $M_{\rm sd} = 0.6\,M_\odot$. This means that the clump will truly be lost to the system. The *total* mass of the clump, calculated as the mass passing the computational boundary during the ejection event, is 0.15 $M_\odot$. It should be noted, however, that this value includes not only the compact core but also a diffuse envelope and even fragments of a spiral. We speculate that upon contraction this clump may form a substellar object, given that a significant fraction of mass remains in the disk until it is ejected due to outflows and/or dispersed due to photoevaporation. No further major fragmentation episodes in the protostellar disk are seen after this event, likely due to a sharp drop in the total disk mass caused by the ejection. The disk then gradually evolves toward a nearly axisymmetric state.

## 7. CRYSTALLIZATION OF SILICATES IN PROTOSTELLAR DISKS

The degree of crystallinity of silicates in prestellar cores is about 0.2% ± 0.2%, similar to that of the interstellar medium (Kemper et al. 2004). On the other hand, the crystalline-silicate fraction in the inner regions of circumstellar disks around young stellar objects (YSOs) can vary in wide limits from essentially none to almost 100% (Watson et al. 2009), implying in situ production of crystalline silicates in at least some YSOs. Two mechanisms for dust crystallization have been put forward: (a) evaporation of the original, amorphous dust grains followed by recondensation under conditions of high temperature and density (e.g., Grossman 1972) and (b) thermal annealing of the amorphous grains at temperatures (800–1300) K via viscous heating (e.g., Gail 2001), shock wave heating (Harker and Desch 2002), or disk surface heating during EX-Lupi-like outbursts (Ábrahám et al. 2009). Both mechanisms are thought to work mostly in the inner several AU from the star. However, recent missions to comets Wild 2 and Tempel 1 have brought evidence for crystalline silicates in these comets (e.g., McKeegan et al. 2006). This suggests that some means of outward radial transport of crystallized dust to distances of the order of tens of AU is necessary to account for the non-zero crystalline-silicate fraction in comets (e.g., Gail 2001). Recently, Vorobyov (2011) has suggested another promising gateway for the production of crystalline silicates in protostellar disks, which does not require outward transport of dust. Crystalline silicates can be produced by thermal annealing of amorphous dust grains in the hot interiors of massive fragments (or planet/brown-dwarf embryos) and released into the disk when the fragments are tidally dispersed. Fragments can be dispersed at various distances from sub-AU to several tens or even hundreds of AU, enriching the disk with crystalline silicates during the first Myr of evolution. These crystalline silicates can later be incorporated into comets as dust sediments and grows to form larger bodies.

Vorobyov modified the model described in Section 2 by adding two additional continuity equations for the surface density of amorphous and crystalline silicates, including the effect of silicate crystallization at temperatures > 800 K:

$$\frac{\partial \Sigma_{a.s.}}{\partial t} + \nabla_p \cdot (\Sigma_{a.s.} \boldsymbol{v}_p) = -S,$$

$$\frac{\partial \Sigma_{c.s.}}{\partial t} + \nabla_p \cdot (\Sigma_{c.s.} \boldsymbol{v}_p) = +S,$$

where $S = \nu_{cr} \Sigma_{a.s.}$ and the crystallization rate $\nu_{cr}$ is provided by the Arrhenius formula (see Vorobyov 2011).

Figure 5 presents the gas surface density (left), gas temperature (middle) and, crystalline silicate fraction $\xi = \Sigma_{c.s.}/(\Sigma_{c.s.} + \Sigma_{a.s.})$ in the inner 400 AU of a young protostellar disk formed from a cloud core with initial mass $M_c = 0.78\ M_\odot$ and $\beta = 8.8*10^{-3}$. There are three distinct fragments present in the disk, but only one (shown by the arrow) is characterized by the gas temperature in its interior exceeding 800 K (middle panel). Such temperatures are sufficient to anneal pristine amorphous silicates so that the fragment now contains mostly crystalline silicates (right panel). Numerical hydrodynamics simulations show that fragments containing thermally processed dust in their interiors can be dispersed by tidal torques at various radial distances from (sub-) AU scales, if they are destroyed on their approach to the central star, to hundred-AU scales, if they are dispersed by tidal torques exerted by spiral arms (Vorobyov 2011). When dispersed, these fragments release crystalline silicates directly into the disk.

We stress that this mechanism can provide a direct source of crystalline silicates at large distances, thus naturally explaining the relatively large abundance of crystalline silicates in such comets as Wild 2 and Tempel 1. The presence of crystalline silicates with the chemical and isotopic composition similar to that of the chondritic meteorites does not invalidate this mechanism, as the latter may have formed in the atmospheres of protoplanetary embryos via dust sedimentation and released into the inner few AU when the embryos were dispersed by tidal torques near the young protosun. Also the presence of refractory calcium aluminum inclusions in the cometary material can be accounted for by the fact that temperatures in at least some of the embryos can exceed 1300 K.

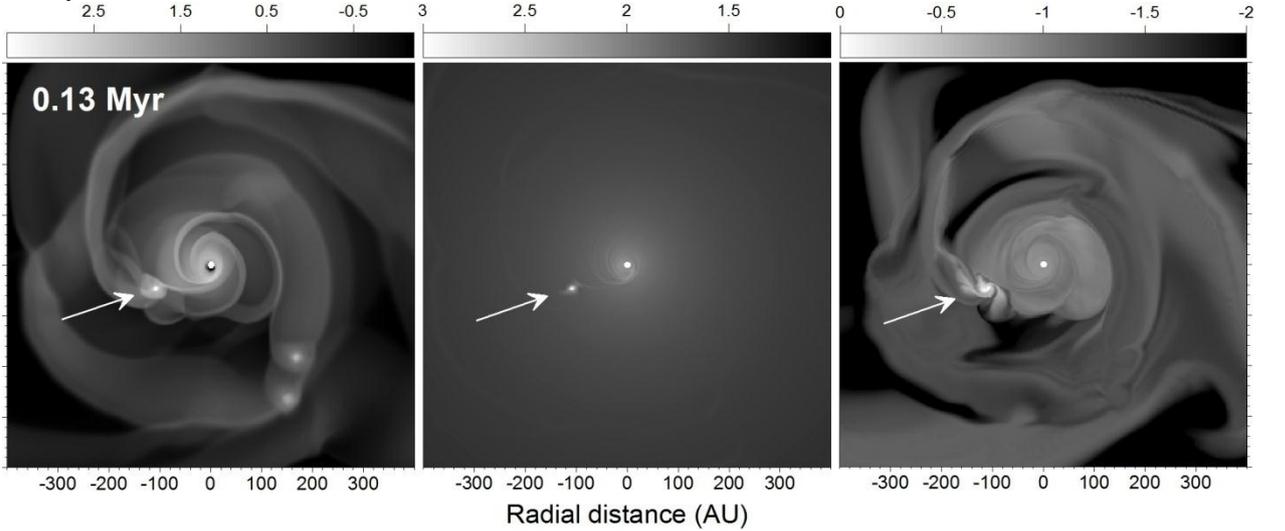

Fig. 5. Gas surface density (left panel, g cm$^{-2}$), gas midplane temperature (middle panel, K), and the crystalline silicate fraction (right panel) in the protostellar disk at $t = 0.13$ Myr after the formation of the central star. Only the inner 300 AU are shown. All quantities are plotted on the log scale. The gas temperature in the depths of the most massive and densest clump (shown by the arrow) reaches temperatures $\geq 800$ K, sufficient to anneal amorphous silicates.

CONCLUSIONS

This article describes the new Migrating Embryo model for disk evolution, which is revealing many new pathways to understanding the formation of stars, brown dwarfs, gas giant planets, and even the processing of solids in the early solar system. We believe that the study of the formation of solids and planets in the early solar system can be moved forward in new directions by examining implications of this new scenario. While the well-established Core Accretion model has been very successful in explaining many features of planet formation, we feel that the Migrating Embryo model provides alternative and complementary pathways for many outcomes, including planet formation and the thermal processing of solids. The implications are just beginning to be explored, and provide a rich terrain for both astronomers and cosmochemists to explore. In the future, we plan to include a more detailed modeling of gas hydrodynamics in our thin-disk code by reconstructing the disk vertical structure using equations of vertical hydrostatic equilibrium and radiation transfer. This modification would allow us to include chemical reaction networks and dust evolution into the resulting 2+1 dimensional numerical hydrodynamics code. Model predictions of dust density and temperature can be used to successfully interface with high-resolution observations of planet-forming disks using the Atacama Large Millimeter/Submillimeter Array and other future observatories.

ACKNOWLEDGEMENTS

We are thankful to the referees for suggestions and comments that helped greatly to improve the manuscript. Numerical simulations were done at the Atlantic Computational Excellence Network (ACEnet), the Shared Hierarchical Academic Research Computing Network (SHARCNET), and the Center of Collective Supercomputer Resources, Taganrog Technological Institute at Southern Federal University. Support for this work was provided by an NSERC Discovery Grant to SB and RFBR grants 10-02-00278 and 11-02-92601 to EIV.